\documentclass[a4paper]{jpconf}
\usepackage{graphicx}
\begin{document}
\pagenumbering{arabic}
\title{SuperNEMO double beta decay experiment}

\author{Alexander Barabash (for the SuperNEMO Collaboration)}

\address{Institute of Theoretical and Experimental Physics, 
B.Cheremushkinskaya 25, 117218 Moscow, Russia}

\ead{barabash@itep.ru}

\begin{abstract}
SuperNEMO is a next-generation double beta decay experiment based on the successful 
tracking plus calorimetry design approach of the recently stopped NEMO3 experiment. SuperNEMO
can study a range of isotopes, but the baseline isotope is $^{82}$Se. The total isotope 
mass will be 100--200 kg. A sensitivity to a $0\nu\beta\beta$ half-life greater than 
$10^{26}$ years can be reached which gives access to Majorana neutrino masses of
50--100 meV. Having successfully completed R\&D stage the SuperNEMO Collaboration 
has commenced the construction of the first module, the Demonstrator. The present status of 
SuperNEMO program and plan for the nearest future are discussed.
\end{abstract}

\section{Introduction}
During $\sim$ 9 years (2002-2010) NEMO3 detector was operated in the Modane underground laboratory (LSM) 
in the Frejus tunnel between France and Italy. Seven different isotopes have been studied in NEMO3 
and very interesting and important results for 2$\nu$- and 0$\nu$-decay modes have been obtained 
for $^{100}$Mo \cite{ARN05,BON11}, $^{82}$Se \cite{ARN05,BON11}, $^{150}$Nd \cite{ARG09}, 
$^{96}$Zr \cite{ARG10}, $^{130}$Te \cite{ARN11}, $^{116}$Cd \cite{BON11}, 
and $^{48}$Ca \cite{BON11}. The tracking plus calorimetry technique employed in NEMO3 
provides an accurate and efficient identification of useful and background evens. And we 
have decided to use this very well known technique for new (SuperNEMO) experiment. 
Main advantages of the NEMO technique are the following:

 - full event reconstruction;

 - clear  event signature;

 - excellent background rejection;

 - new physics studies using event topology (mass mechanism, 
    RHC, excited states,...) \cite{ARN10}.
In addition we plan to use planar geometry (instead cylindrical in NEMO3), to use modular 
system and improve some characteristics of the detector (energy resolution, efficiency, 
purity of the source, etc.). This report will focus on the present status of 
SuperNEMO experiment, main recent achievements and plans for the nearest future.

\section{SuperNEMO}
SuperNEMO aims to extend and improve the successful NEMO3 technology. It will extrapolate 
NEMO3 by one order of magnitude by studying about 100-200 kg of  $\beta\beta$ isotope(s). 
The detector's ability to measure any $\beta\beta$ isotope and reconstruction of the 
topological signature of the decay are distinct features of SuperNEMO. The baseline 
isotope choice for SuperNEMO is $^{82}$Se. However other isotopes are possible. 
In particular, $^{150}$Nd and $^{48}$Ca are being looked at. Detector will be able 
to measure individual electron tracks, vertices, energies and time of flight, and 
to reconstruct fully the kinematics and topology of an event. Particle identification 
of gamma and alpha particles, as well as distinguishing electrons from positrons with 
the help of a magnetic field, form the basis of background rejection. An important 
feature of NEMO3 which is kept in SuperNEMO is the fact that the double beta decay 
source is separate from the detector, allowing several different isotopes to be studied. 
SuperNEMO will consist of about twenty identical modules, each housing around 
5–-7 kg of isotope. The project is completing a 3 year design study and R\&D phase 
with much progress towards the first prototype Demonstrator module. The R\&D program 
focuses on four main areas of study: isotope enrichment, tracking detector, calorimeter, 
and ultra-low background materials production and measurements. The expected improvement 
in performance of SuperNEMO compared to its predecessor NEMO3 is shown in Table 1 which 
compares the parameters of the two experiments. The most important design study tasks 
are described in the sections that follow.



\begin{center}
\begin{table}[h]
\caption{\label{tab:table1}Comparison of main NEMO3 and SuperNEMO parameters.}
\centering
\begin{tabular}{lll}
\br
& NEMO3 & SuperNEMO\\
\mr
Isotope & $^{100}$Mo & $^{82}$Se \\
mass& 7 kg & 100--200 kg\\
signal efficiency & 18\% & $>$ 30\% \\
$^{208}$Tl in foil & $<$ 20 $\mu$Bq/kg & $<$ 2 $\mu$Bq/kg \\
$^{214}$Bi in foil & $<$ 300 $\mu$Bq/kg & $<$ 10 $\mu$Bq/kg \\
energy resolution at 3 MeV & 8\% (FWHM) & 4\% (FWHM) \\
Sensitivity to half-life & $\sim$ 10$^{24}$ yr & $\sim$ $(1-2)\times 10^{26}$ yr \\
sensitivity to neutrino mass & $\sim$ (0.3-0.9) eV & $\sim$ (0.05--0.1) eV \\
\br
\end{tabular}
\end{table}
\end{center}

Figure 1 shows two renderings of the SuperNEMO Demonstrator module. The source is a thin 
($\sim$ 40 mg/cm) foil
inside the detector. It is surrounded by a gas tracking chamber followed by calorimeter walls. 
The tracking volume
contains more than 2000 wire drift chambers operated in Geiger mode, which are arranged in 
nine layers parallel to
the foil. The calorimeter is divided into $\sim$ 1000 blocks which cover most of the detector 
outer area and are read out by
photo multiplier tubes (PMT).

\begin{figure}[h]
\begin{minipage}{14pc}
\hspace{5pc}
\includegraphics[width=11pc]{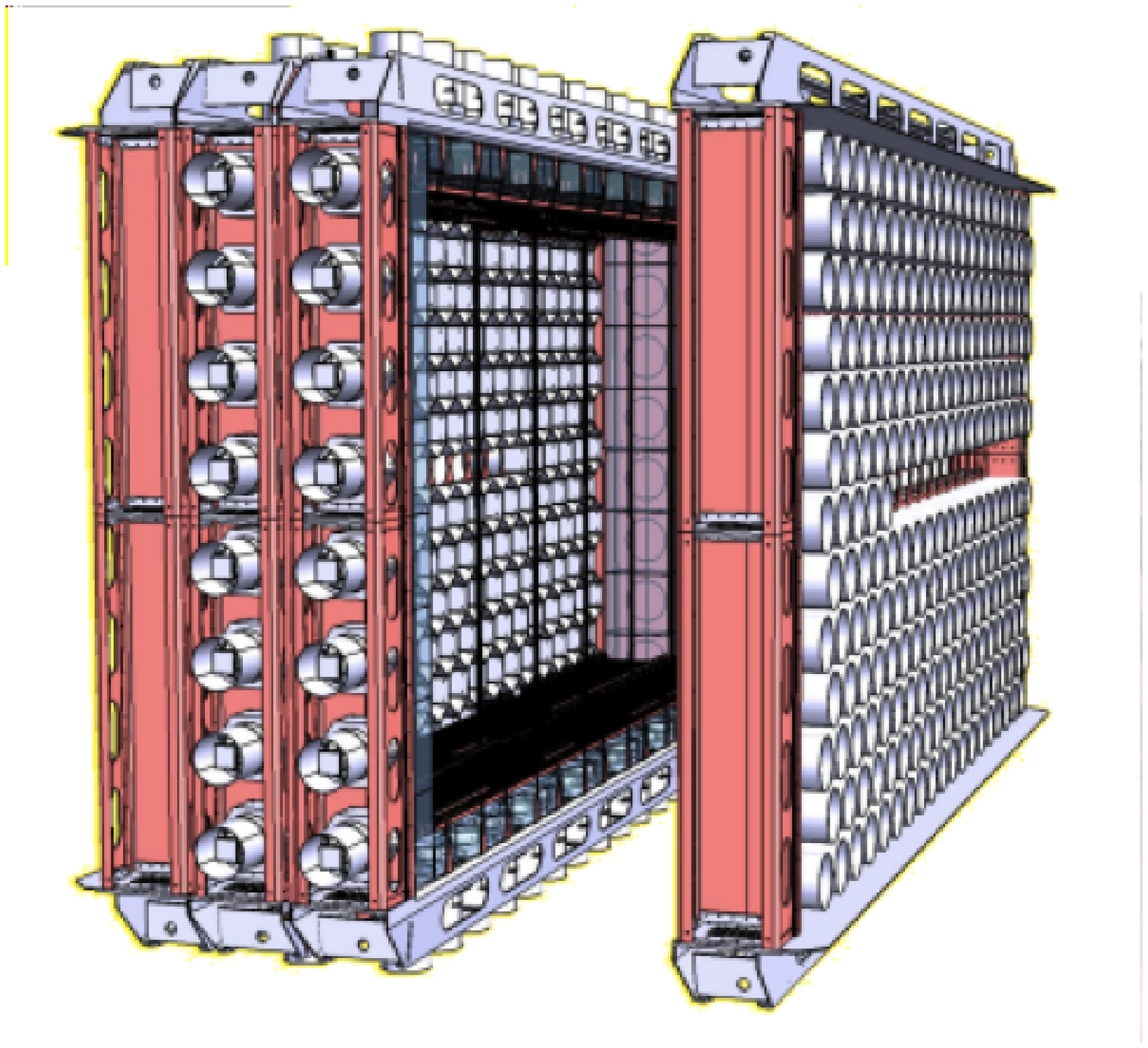}
\end{minipage}\hspace{5pc}%
\begin{minipage}{14pc}
\includegraphics[width=14pc]{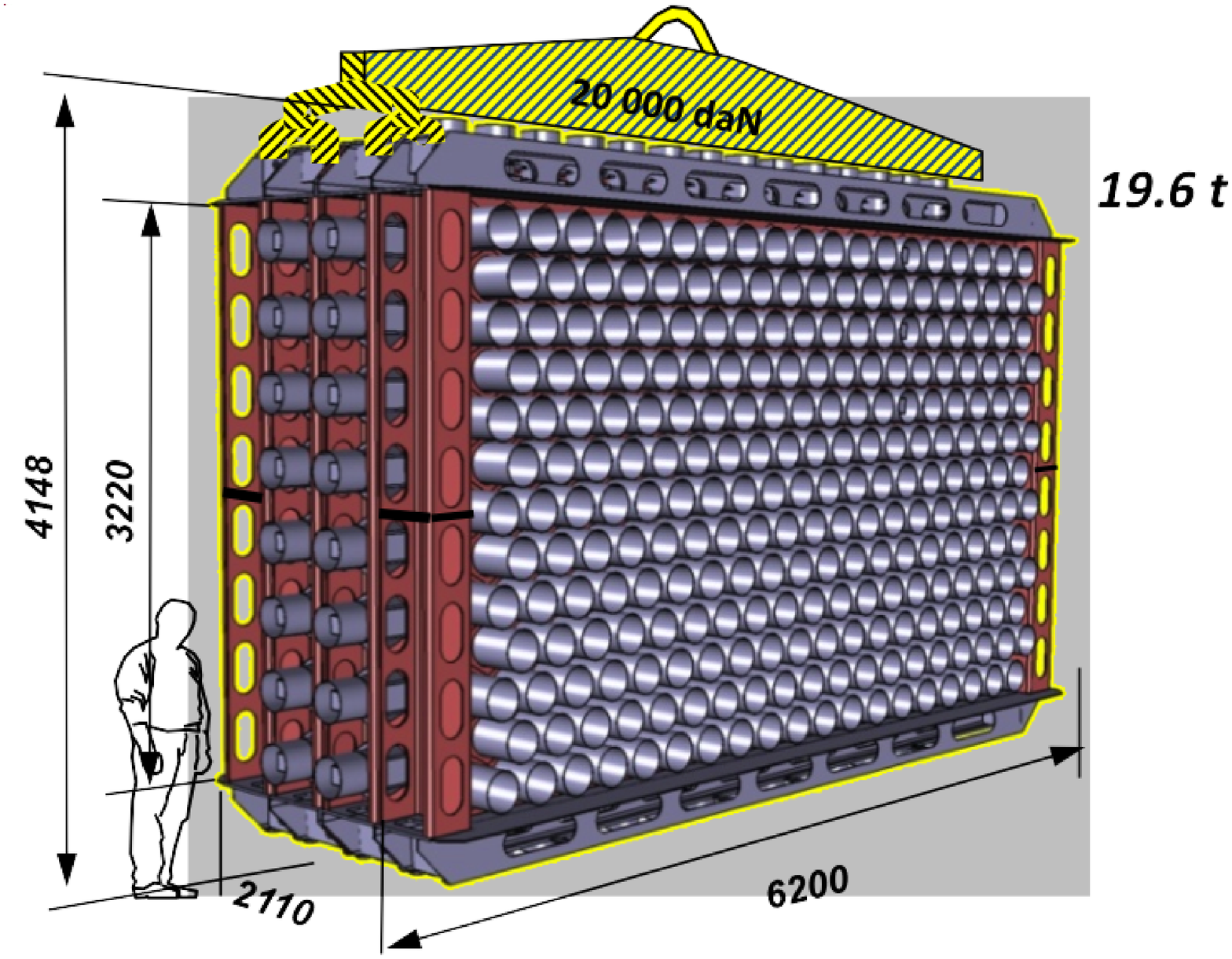}
\end{minipage}
\caption{\label{label}Preliminary design of the SuperNEMO detector. Left-hand side:
An exploded view showing the tracking chamber and calorimeter modules. Right-hand side: 
A view showing the configuration of the calorimeter modules with dimensions.}
\end{figure}

\subsection{Calorimeter R\&D}

SuperNEMO aims to improve the calorimeter energy resolution to 7\%$/E$ at FWHM (4\% at 
the Q$_{\beta\beta}$ energy). To reach this goal, several studies have been completed 
on the choice of calorimeter parameters such as scintillator material (organic plastic 
or liquid), and the shape, size and coating of calorimeter blocks. These are combined 
with dedicated development of PMTs with low radioactivity and high quantum efficiency.
The feasibility to reach the required resolution of 7--8\%/$\surd$E (MeV) with a large 
(26x26x15 cm$^3$) block has been experimentally demonstrated with PVT-based scintillators 
coupled to a low radioactive high-QE 8" R5912MOD Hamamatsu PMT.
The remaining challenge is to demonstrate that the achieved energy 
resolution can be maintained at the mass production scale. The large scale construction will 
start in 2012. 

\subsection{Tracker design}

The SuperNEMO tracker consists of octagonal wire drift cells operated in Geiger mode. Each 
cell is around 4 m long and has a central anode wire surrounded by 8--12 ground wires, with 
cathode pickup rings at both ends. Signals can be read out from the anode and/or cathodes to 
determine the position at which the ionizing particle crossed the cell. The tracking detector 
design study looks at optimizing its parameters to obtain high efficiency and resolution in
measuring the trajectories of double beta decay electrons, as well as of alpha particles for 
the purpose of background rejection. The tracking chamber geometry is being investigated with 
the help of detector simulations to compare the different possible layouts. In addition, several 
small prototypes have been built to study the drift chamber cell design and size, wire length, 
diameter and material, and gas mixture \cite{NAS07}. The first 9-cell prototype was successfully operated 
with three different wire layouts, demonstrating a plasma propagation efficiency close to 100\% 
over a wide range of voltages \cite{NAS07}. In addition, a 90-cell prototype has recently been constructed. 
Measurements of cosmic ray tracks in a 90-cells prototype have been demonstrated the required space resolution 
(0.7 mm radial and 1 cm longitudinal). A SuperNEMO
module will contain several thousand drift cells with 8--12 wires each. The large total number 
of wires requires an automated wiring procedure, thus a dedicated wiring robot is being developed 
for the mass production of drift cells.

\subsection{Choice of source isotope}

The choice of isotope for SuperNEMO is aimed at maximizing the neutrinoless signal over the 
background of two-neutrino double beta decay and other background events. Therefore the isotope 
must have a long two-neutrino half-life, and high endpoint energy and phase space factor. The 
enrichment possibility on a large scale is also a factor in selecting the isotope. The main 
candidate isotopes for SuperNEMO have emerged to be $^{82}$Se (E$_{2\beta}$ = 2995 keV). 
A sample of 5 kg of $^{82}$Se 
has been enriched and is currently undergoing purification.

\subsection{Radiopurity of the source}

SuperNEMO will search for a very rare process, therefore it must maintain ultra-low background levels. 
The source foils must be radiopure, and their contamination with naturally radioactive elements must 
be precisely measured. The most important source contaminants are $^{208}$Tl and $^{214}$Bi, whose 
decay energies are close to the neutrinoless double beta decay signal region. SuperNEMO requires 
source foil contamination to be less than 2 $\mu$Bq/kg for $^{208}$Tl and less than 10 $\mu$Bq/kg for 
$^{214}$Bi.
In order to evaluate these activities, a dedicated "BiPo" detector was developed which can measure 
the signature of an electron followed by a delayed alpha particle. The first BiPo prototype (BiPo1) 
 was installed in the Modane Underground Laboratory in February 
2008 and is currently running with 20 modules (with a 0.8 m$^2$ of sensitive surface). 
The objective for this prototype is to measure the backgrounds and surface contamination of the 
prototype's plastic scintillators \cite{BiPo10}. 
A medium-size BiPo-3 detector with a 3.25 m$^2$ sensitive surface and using the same techniques developed in 
the BiPo-1 prototype is under construction (see Fig. 2). The goal of the BiPo-3 detector is to 
measure the first double beta 
decay source foils of the SuperNEMO Demonstrator in the year 2012. Required SuperNEMO sensitivity 
will be reached in six months of measurement.

\begin{figure}[h]
\begin{center}
\includegraphics[width=16pc]{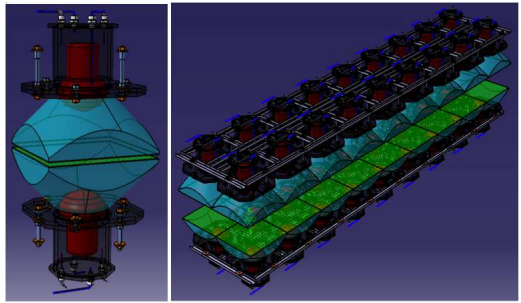}
\end{center}
\caption{\label{label}Schematic view of BiPo-3 capsule (left) and one of two modules (right).}
\end{figure}

\section{SuperNEMO Demonstrator and schedule outline}

Having successfully completed the R\&D stage the
SuperNEMO collaboration has commenced the construction
of the first module, the Demonstrator. The
main goals of the Demonstrator are: to demonstrate feasibility
of mass production of detector components under
ultra-low background conditions; to measure backgrounds
especially from radon emanation; to finalize the
detector design; to produce a competitive physics
result. To accomplish the latter goal on a competitive
time scale the Demonstrator module will house 5--7 kg of
the $^{82}$Se isotope. The construction and commissioning
of the Demonstrator will be completed in 2013 with
data taking expected to commence in the second half
of 2013. The module will be located in the LSM (in the existing cavern).
 The sensitivity of
the Demonstrator after 17 kg$\times$yr of exposure is 6.5$\times$10$^{24}$
yr (90\% CL) (sensitivity to effective Majorana neutrino mass is $\sim$ 0.2-0.5 eV). 
The modular design of SuperNEMO makes it possible
to proceed with construction and data taking in parallel.
The full detector construction is expected to start
in 2014 (in parallel with the Demonstrator running). The
500 kg$\times$yr exposure will be reached in $\sim$ 2019 pushing the
sensitivity to the effective Majorana neutrino mass down
to 50–-100 meV.

\section{Conclution}

Based on the successful experience of the NEMO detectors, the extensive and
intensive SuperNEMO R\&D program is finishing with construction of the 
Demonstrator started in 2010. Construction and commissioning of the first 
module (Demonstrator) will be completed in 2013 for a possible full detector 
construction starting in 2014 reaching its sensitivity of $\sim$ 10$^{26}$ yr on 
$0\nu\beta\beta$ decay of $^{82}$Se in $\sim$ 2019. The unique technique of the SuperNEMO 
detector could provide the possibility to study the origin of $0\nu\beta\beta$ decay
in the case of its discovery.

\end{document}